\documentclass[12pt]{article}
\usepackage{amsmath,cite,epsfig,a4,times,euscript,listings}
\usepackage[text-hat-accent]{euler}
\usepackage{graphics}
\usepackage[usenames]{color}
\usepackage[T1]{fontenc}
\usepackage{bold-extra}
\usepackage[nottoc,numbib]{tocbibind}
\renewcommand{\baselinestretch}{1.1}
\newcommand{\myTitle}[1]{\begin{center}{\bf\Huge #1}\\[5ex]\end{center}}
\newcommand{\myAuthor}[1]{\begin{center}{\Large #1}\\[2ex]\end{center}}
\newcommand{\myAffiliation}[1]{\\[1ex]{\it\large #1}}
\newcommand{\myEmail}[1]{}
\newcommand{\myDate}{\vspace{5ex}}
\newcommand{\myAbstract}[1]{\begin{center}\renewcommand{\baselinestretch}{1}{\bf Abstract}\\[2ex]\parbox{0.8\linewidth}{\small\hspace{15pt} #1}\end{center}\vspace{\baselineskip}}
\newcommand{\myReport}[1]{\hspace{\fill} #1}
\newcommand{\myPreprint}[1]{}
\newcommand{\myKeywords}[1]{}
\newcommand{\myFigure}[1]{\begin{figure}\begin{center}#1\end{center}\end{figure}}

\newcommand{\slashp}{p\hspace{-6.5pt}/}

\newcommand{\slashk}{k\hspace{-6.5pt}/}

\newcommand{\slasheps}{\varepsilon\hspace{-6.0pt}/}

\newcommand{\Figure}[1]{Fig.~\ref{#1}}

\newcommand{\KaTie}{{\sc Ka\nolinebreak\hspace{-0.3ex}Tie}}

\newcommand{\imag}{\mathrm{i}}

\newcommand{\lop}[2]{#1\!\cdot\!#2}

\newcommand{\tweakcodepar}[3]%
  {\vspace{#1ex}\newline\noindent\hspace*{4.0ex}{\small\tt #3}\vspace{#2ex}\newline\noindent}

\begin{document}

\myReport{IFJPAN-IV-2019-2}
\myPreprint{}\\[2ex]

\myTitle{%
A note on QED gauge invariance\\[0.5ex]
of off-shell amplitudes
}

\myAuthor{%
A.~van~Hameren%
\myAffiliation{%
Institute of Nuclear Physics Polish Academy of Sciences\\
PL-31-342 Krak\'ow, Poland%
\myEmail{hameren@ifj.edu.pl}
}
}

\myDate

\myAbstract{%
The consistent construction of scattering amplitudes for processes with off-shell initial partons and involving electro-weak interactions is addressed.
}

\myKeywords{QCD, QED, off-shell, tree-level, Standard Model}

%
In the calculation of a cross section for the scattering of hadrons one employs a factorization prescription, separating the relatively low-energy initial-state interactions from a partonic hard scattering.
Some factorization prescriptions explicitly involve non-vanishing momentum components for the initial-state partons that are transverse to the direction of the hadrons, rendering them off-shell.
The amplitudes relevant for the partonic process are referred to as {\em off-shell amplitudes\/}, and some care has to be taken to ensure that they satisfy gauge invariance with respect to quantum chromodynamica (QCD)~\cite{Lipatov:1995pn,Lipatov:2000se,vanHameren:2012if,vanHameren:2013csa,Kotko:2014aba}.

Recently, it has been highlighted that gauge invariance for amplitudes with off-shell partons is also a non-trivial issue with respect to quantum electrodynamica (QED) if the off-shell initial-state partons are quarks~\cite{Nefedov:2018vyt}.  
In that paper, an effective quark-antiquark-photon vertex attributed to Fadin and Sherman~\cite{Fadin:1976nw} is the essential ingredient to guarantee QED gauge invariance.  
I explain in this note how QED gauge invariance is ensured within the method of~\cite{vanHameren:2013csa}.

The approach to off-shell amplitudes of ~\cite{vanHameren:2012if,vanHameren:2013csa} ensures QCD gauge invariance by replacing each off-shell parton by a pair of auxiliary on-shell partons satisfying eikonal Feynman rules.
Each flavor gets its own auxiliary partons.
Let us denote the auxiliary quark associated with an off-shell gluon by $q_A$, and the auxiliary quarks associated with the flavors $u,d,\ldots$ etcetera by $u_A,d_A,\ldots$ etcetera.
The anti-quarks $\bar{q}_A,\bar{u}_A,\bar{d}_A,\ldots$ are also needed.
Off-shell intial-state quarks finally also require an auxiliary photon denoted by $\gamma_A$.
A partonic process involving an off-shell initial-state gluon, say 
%
\begin{equation}
g^*\,y_1\to y_2\,y_3
\quad\textrm{is replaced with}\quad
q_A\,y_1\to q_A\,y_2\,y_3
\quad,
\end{equation}
%
where $y_1,y_2,y_3$ indicate some other particles and partons involved in the particular process.
For an off-shell (anti-)quark, for example
%
\begin{align}
u^*\,y_1\to u\,y_3
&\quad\textrm{is replaced with}\quad
u_A\,y_1\to \gamma_A\,u\,y_3
\quad,
\\
\bar{u}^*\,y_1\to \bar{u}\,y_3
&\quad\textrm{is replaced with}\quad
\bar{u}_A\,y_1\to \gamma_A\,\bar{u}\,y_3
\quad.
\end{align}

We need to distinguish the combinatorial content from the kinematical content of the Feynman rules.
In the most practical formulation, the latter only affects the propagators of internal auxiliary quark lines, and spinors and polarization vectors of external auxiliary partons, while vertices are not affected~\cite{Hameren:2014qza}.
Practicality refers to bookkeeping here; computationally a price is payed by keeping the spinorial nature of auxiliary quark lines rather than turning them into scalar lines. 
The rules relevant for the current discussion are completely specified as follows: external auxiliary (anti-)quarks are associated with a massless spinor defined with the light-like longitudinal momentum $p_A^\mu$ of the initial-state hadron, and the auxiliary photon is associated with a polarization vector defined with the same momentum.
Internal auxiliary quark lines are associated with a propagator 
%
\begin{equation}
\frac{\imag\slashp_A}{2\lop{p_A}{K}}
\quad,
\end{equation}
%
where $K^\mu$ is the momentum flowing through the line.
All vertices involving the auxiliary partons are the same as for ordinary partons.
The same holds for the color content.
The auxiliary photon is a color singlet, and the auxiliary (anti-)quarks are in the (anti-)fundamental representation.
Thanks to the eikonal Feynman rules the $q_A\,\bar{q}_A$ pair effectively is a color octet, that is the trace of the amplitude over the color indices of the pair vanishes.
Finally, there is a freedom in the choice of momentum flow (fractions of $p_A^\mu$ and the transverse momentum are eliminated in the auxiliary quark propatagor), but it is most conveniently chosen such that the initial-state auxiliary parton carries the whole off-shell momentum, while the final-state auxiliary parton carries none.

The combinatorial rules tell us which particles couple with each other.
In order to achieve QCD gauge invariance for off-shell gluons, the following coupling must be included:
%
\begin{equation}
q_A\,\bar{q}_A\,g
\quad.
\end{equation}
%
In order to allow for off-shell (anti-)quarks, the following couplings must be included for each flavor:
%
\begin{equation}
u_A\,\gamma_A\,\bar{u}
\quad,\quad
\bar{u}_A\,\gamma_A\,u
\quad,\quad
u_A\,\bar{u}_A\,g
\quad.
\end{equation}
%
If both inital-state partons are off-shell, then the set of auxiliary partons must be duplicated to include $q_B,u_B,d_B,\ldots$ etcetera.
No interactions involving both $A$-type and $B$-type auxiliary partons are introduced.
Notice that this implies that there cannot be internal auxiliary photon lines. 

QED gauge invariance is guaranteed by including the coupling of the auxiliary quark partons to electro-weak vector bosons:
%
\begin{gather}
u_A\,\bar{u}_A\,\gamma
\;\;,\;\;
u_A\,\bar{u}_A\,Z
\;\;,\;\;
d_A\,\bar{d}_A\,\gamma
\;\;,\;\;
d_A\,\bar{d}_A\,Z
\;\;,\;\;
u_A\,\bar{d}_A\,W^-
\;\;,\;\;
\bar{u}_A\,d_A\,W^+
\;\;,\;\;
\end{gather}
%
and the same for the other families.
Notice that no interaction of the electro-weak vector bosons with the auxiliary gluon (anti-)quark $q_A,\bar{q}_A$ are included.
Again, the vertices are just those from standard electro-weak theory.

\Figure{fig:uguA} shows an example for the process $u^*\,g^*\to u\,\gamma$.
\myFigure{%
\epsfig{figure=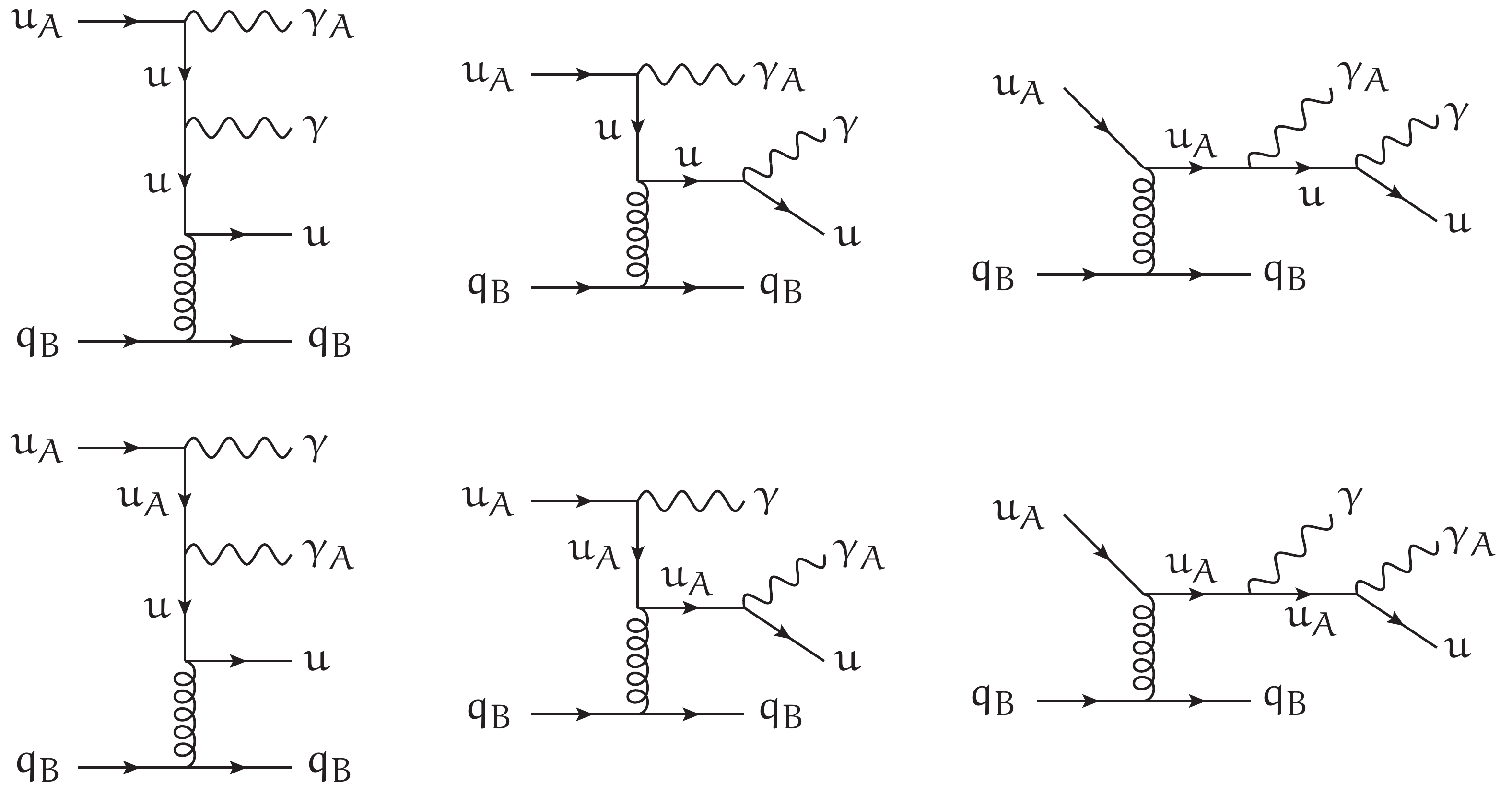,width=0.72\linewidth}
\caption{\label{fig:uguA}%
Feynman graphs for the process $u^*\,g^*\to u\,\gamma$. The graphs on the second row contain the extra $u_A\,\bar{u}_A\,\gamma$ vertices in order to ensure QED gauge invariance.
}
}
The first two graphs in the top row have the on-shell topology, and the third one on the top row ensures QCD gauge invariance.
The second row must be included to also ensure QED gauge invariance.
Notice that the photon $\gamma$ does not couple to the auxiliary gluon quark $q_B$.

\Figure{fig:uuWW} shows an example for the process $u^*\,\bar{u}^*\to W^+W^-$.
\myFigure{%
\epsfig{figure=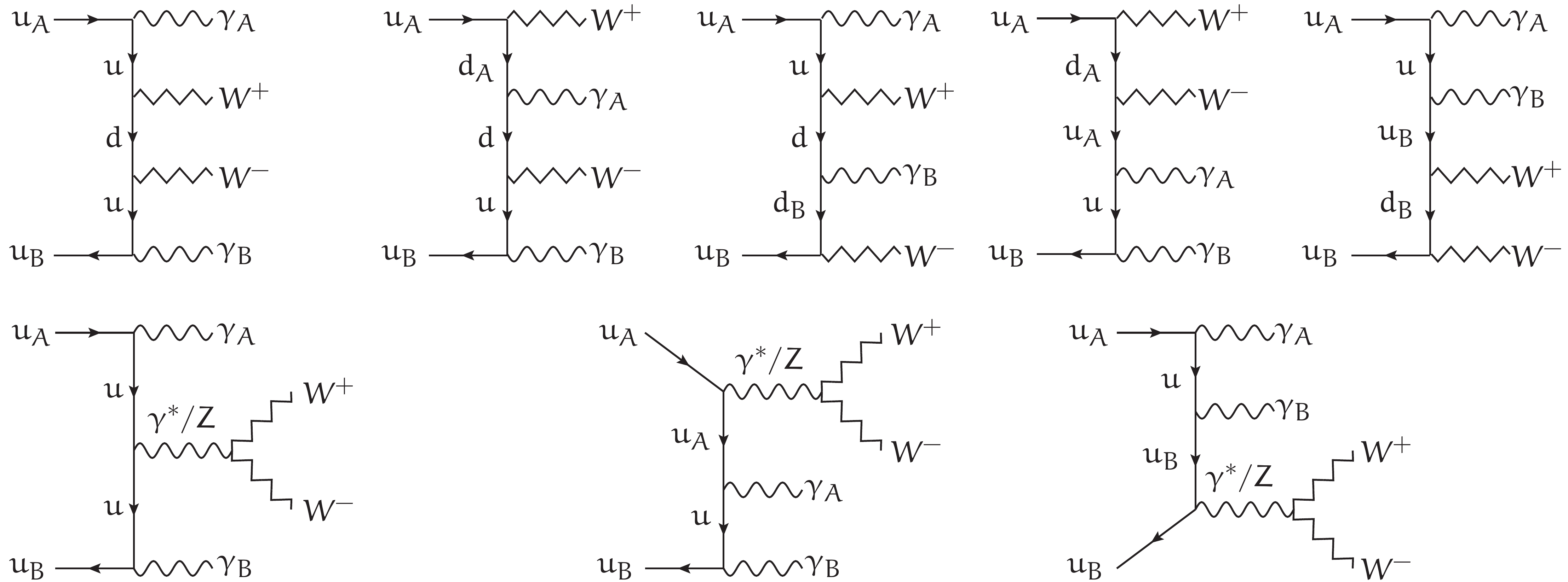,width=\linewidth}
\caption{\label{fig:uuWW}%
Feynman graphs for the process $u^*\,\bar{u}^*\to W^+W^-$. For clarity, flavors are indicated without referring to anti-quarks, so as if auxiliary photon $\gamma_B$ instead of the auxiliary quark $u_B$ were in the initial state.
}
}
The two graphs on the left are those that would be included without the extra interaction, and on their own would break QED gauge invariance.
One can easily see that the three lower graphs together compose the Fadin-Sherman effective vertex.
Denoting the momenta of the off-shell partons with $k_A^\mu,k_B^\mu$, and omitting common overall factors, the three graphs read
%
\begin{equation}
\bar{u}_B\,\slasheps_B\,\frac{\slashk_B}{k_B^2}\,\Omega^\mu\,\frac{\slashk_A}{k_A^2}\,\slasheps_A\,u_A
\;-\;
\bar{u}_B\,\slasheps_B\,\frac{\slashk_B}{k_B^2}\,\slasheps_A\,\frac{\slashp_A}{2\lop{p_A}{k_B}}\,\Omega^\mu\,u_A
\;-\;
\bar{u}_B\,\Omega^\mu\,\frac{\slashp_B}{2\lop{p_B}{k_A}}\,\slasheps_B\,\frac{\slashk_A}{k_A^2}\,\slasheps_A\,u_A
\;\;.
\label{Eq:FadinSherman}
\end{equation}
%
The relative minus signs come from momentum flow, and can be understood by replacing $k_B^\mu$ with $-k_B^\mu$.
The symbol 
%
\begin{equation}
\Omega^\mu = \omega\gamma^\mu
\quad,\quad
\omega=v-a\gamma^5
\end{equation}
%
represents the general electro-weak vertex.
Using $\slasheps_A\slashp_A\omega=\omega\slasheps_A\slashp_A$ and the Dirac equation $\slashp_Au_A=0$%
, we get
%
\begin{equation}
\slasheps_A\,\frac{\slashp_A}{2\lop{p_A}{k_B}}\,\omega\gamma^\mu\,u_A
=
\frac{p_A^\mu}{\lop{p_A}{k_B}}\,\omega\,\slasheps_A\,u_A
=
\frac{p_A^\mu}{\lop{p_A}{k_B}}\,\omega\,\slashk_A\,\frac{\slashk_A}{k_A^2}\,\slasheps_A\,u_A
\;\;.
\end{equation}
%
Performing similar operations on the third graph, we find that (\ref{Eq:FadinSherman}) reduces to
%
\begin{equation}
\bar{u}_B\,\slasheps_B\,\frac{\slashk_B}{k_B^2}\,\omega\left[
\gamma^\mu
- \frac{p_A^\mu}{\lop{p_A}{k_B}}\,\slashk_A
- \frac{p_B^\mu}{\lop{p_B}{k_A}}\,\slashk_B
\right]\frac{\slashk_A}{k_A^2}\,\slasheps_A\,u_A
\;\;,
\end{equation}
%
clearly exposing the effective vertex.

All necessary interactions have been implemented into the Monte Carlo program \KaTie~\cite{vanHameren:2016kkz}, and Ward identities of several processes have been tested.
For example for the process $u^*\,\bar{u}^*\to W^+W^-\gamma$ it was numerically confirmed that the amplitude vanishes if the polarization vector of the photon is replaced by its momentum, while it does not vanish if the extra interactions are not included.

\subsection*{Acknowledgments}
All graphs were created with the help of Jaxodraw.
This work was supported by grant of National Science Center, Poland, No.\ 2015/17/B/ST2/01838.

\providecommand{\href}[2]{#2}\begingroup\raggedright\endgroup

\begin{appendix}
\end{appendix}

\end{document}